\documentclass[10pt,sigconf]{acmart}
\usepackage[english]{babel}
\usepackage{graphicx}
\usepackage[caption=false]{subfig}

\renewcommand\footnotetextcopyrightpermission[1]{} % removes footnote with conference info
\setcopyright{none}
\acmDOI{}

\acmISBN{}

\acmPrice{}

\begin{document}
\title{CommunityWatch: The Swiss-Army Knife of BGP Anomaly Detection }

%\titlenote{Produces the permission block, and copyright information}
%\subtitle{Extended Abstract}

\author{Vasileios Giotsas}
% \author{Firstname Lastname}
% \authornote{Note}
 \affiliation{%
   \institution{Lancaster University}
%   \streetaddress{Address}
%   \city{City} 
%   \state{State} 
%   \postcode{Zipcode}
 }
 \email{v.giotsas@lancaster.ac.uk}


%
%\begin{abstract}
%    We present \textit{CommunityWatch}, an open-source system that enables timely and accurate detection of BGP routing anomalies. CommunityWatch leverages meta-data encoded by AS operators on their advertised routes through the BGP Communities attribute. 
%    The BGP Communities values lack standardized semantics, offering the flexibility to attach a wide range of information, including AS relationships, location data, and route redistribution policies. 
%    Therefore, parsing and correlating Community values and their dynamics enables the detection and tracking of a variety of routing anomalies. 
%    We exhibit the efficacy of CommunityWatch through the detection of three different types of anomalies: infrastructure outages, route leaks, and traffic blackholing.
%\end{abstract}

\maketitle

\section{Motivation}

The design of the Internet as a network of independent networks, or Autonomous Systems (ASes), allowed it to spontaneously evolve to  the core communications technology for contemporary society, but also resulted in the ossification of its core protocols, and the opacity of its structure. The current version of BGP~\cite{rfc1654}, the de-facto inter-domain routing protocol, is over two decades old, and despite various revisions since then, a number of serious problems have been building over time, including weak security, non-deterministic behavior, and proneness to misconfiguration~\cite{goldberg2014taking}. 

While the vulnerabilities inherent in the Internet’s architecture have been known for decades, and there has been a great extent of research to address them~\cite{Mitseva2018}, the proposed solutions have not been widely deployed due to the costs and risks involved in replacing the existing network equipment~\cite{Handley:2004}. As a result, most networks rely on reactive defense mechanisms~\cite{zhang2007practical}. Nonetheless, the highly distributed ownership of the Internet infrastructure and its highly dynamic nature make the development of the appropriate anomaly detection mechanisms far from trivial. Operators have full control over their own infrastructure, but little knowledge of what happens beyond their network perimeter. Third-party services can extend the detection capabilities for some classes of anomalies beyond an AS's domain, but the costs involved and concerns with data sharing make many operators reluctant to outsource such functionalities~\cite{sermpezis2018survey}. 
Instad, operators often resort in social media and mailing lists in an effort to crowdsource the debugging of their routing issues~\cite{banerjee2015internet}, an approach that can be error-prone and inefficient.

We take steps toward remedying this situation by developing \textit{CommunityWatch}, an open-source system that enables timely and accurate detection of BGP routing anomalies, by leveraging meta-data encoded by AS operators directly on their BGP messages through the use of the BGP Communities attribute. 

\vspace{-1.35em}
\section{How C{\small ommunity}W{\small atch} Works}

The key insight of our approach is that BGP is no longer purely an “information hiding protocol”~\cite{Ten-Lessons-from-Ten-Years-AS-Modeling:JSAC2011}.
The flattening of the Internet hierarchy~\cite{gill2008flattening}, has led to very dynamic and continuously growing peering clusters, and complex peering practices~\cite{CoNEXT2015-P2F}. Consequently, operators require increasing flexibility and expressiveness in defining their routing policies and communicating them to their neighbors.
The optional BGP Communities attribute~\cite{rfc1997} offers this flexibility by allowing operators to encode arbitrary information on their prefix announcements, including business relationship types, route redistribution policies, location data, and traffic blackholing requests to mitigate attacks~\cite{BGP-Communities:CCR2008}. Their use has become increasingly popular, allowing us to use them as an automated crowdsourcing mechanism for acquiring accurate operator-provided information for about 50\% of IPv4 and 30\% of IPv6 updates. Between 2010 and 2016, the visible ASes using BGP Communities more than doubled, and the number of unique community values tripled to more than 50,000. 

BGP Communities have the format X:Y, where X, Y are two 16-bit
values (extended communities use four octets~\cite{rfc4360}). By convention, the first two octets encode the ASN of the operator that sets the community, while the next two octets encode denote the specific information carried by the Community, as the ingress location of a route. 
Importantly, Communities is a \textit{transitive} attribute,
which means that they can be propagated through multiple AS hops, and we can mine their values through publicly available BGP collectors.

\begin{figure}[t]
	\begin{center}
		\includegraphics[clip,width=0.8\columnwidth]{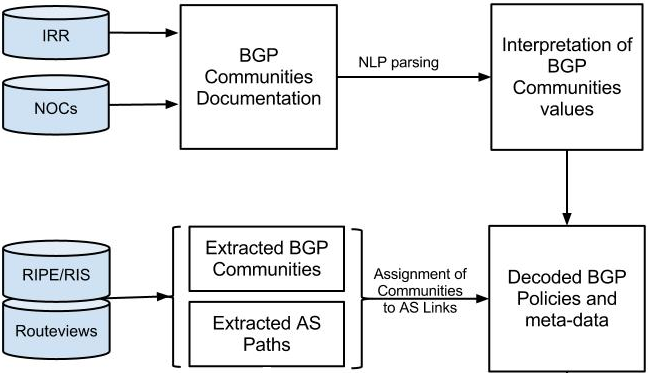}
		\vspace{-1em}
	\end{center}
	\caption{Data collection methodology.}
	\label{fig:communities-methodology}
			\vspace{-1em}
\end{figure}

\begin{figure*}[t]
	\centering
	\subfloat[Aggregated routing activity.]{%
		\includegraphics[clip,width=0.65\columnwidth]{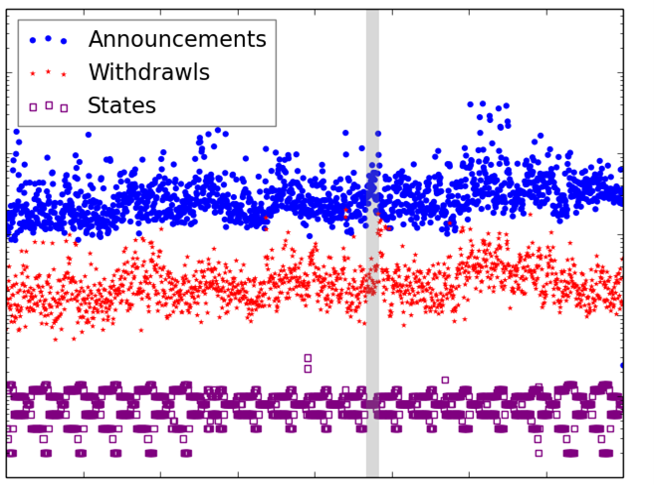}%
		\label{fig:franceix-outage-unfiltered}
	}
	\hfill
	\subfloat[Routing activity filtered based on BGP Communities.]{%
		\includegraphics[clip,width=0.65\columnwidth]{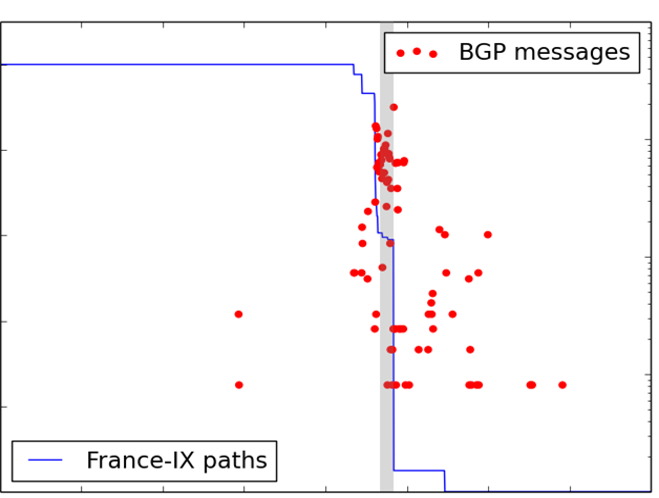}%
		\label{fig:franceix-outage-filtered}
	}
	\hfill
	\subfloat[Longitudinal growth of blackholed prefixes]{%
		\includegraphics[width=0.65\columnwidth]{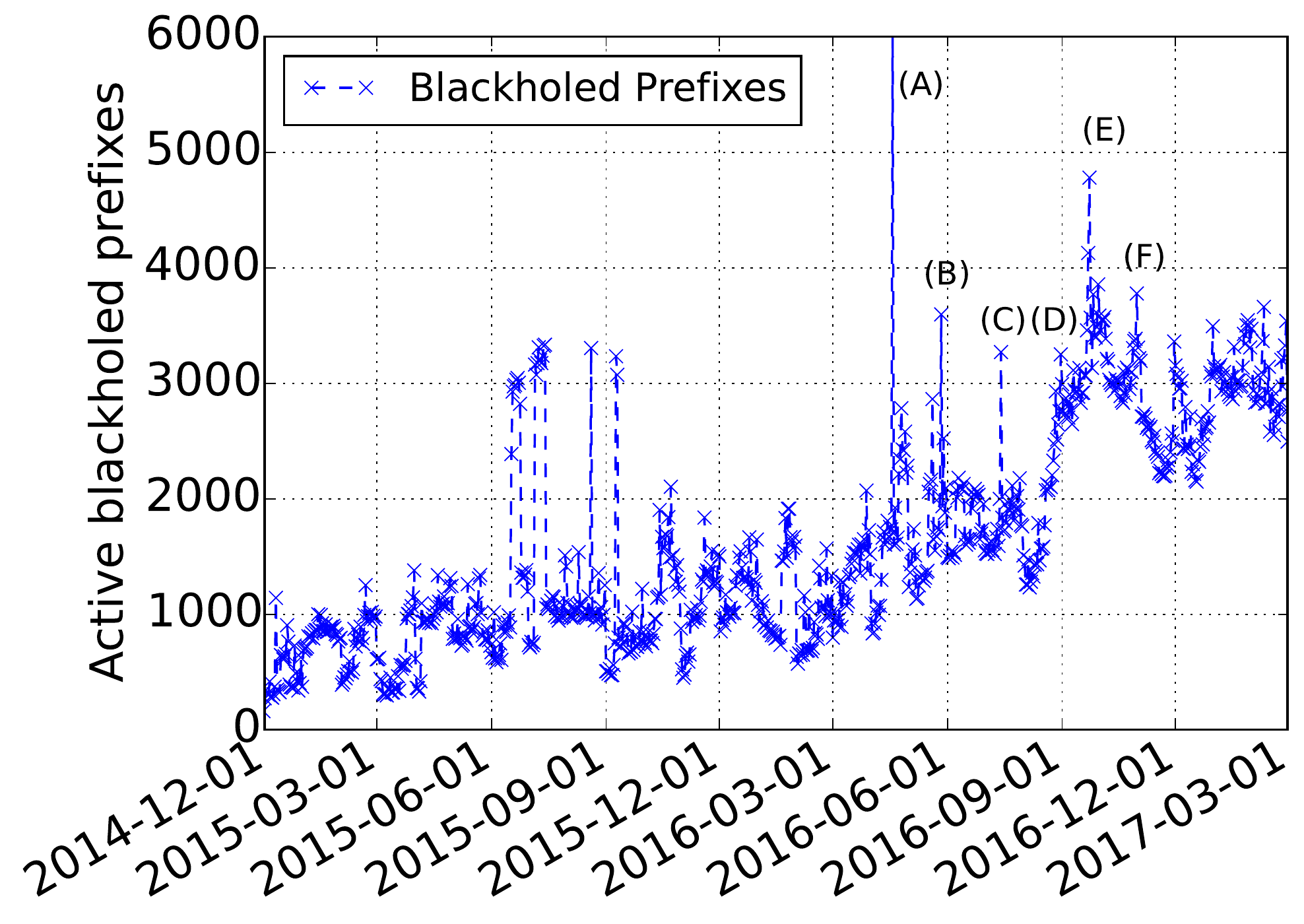}%
		\label{fig:blackholing-growth}
	}
	\vspace{-1em}
	\caption{}
	\label{fig:tuning-and-case-study}
	\vspace{-1.5em}
\end{figure*}

Successful interpretation of the attached Communities values allows monitoring of BGP routes and gathering of routing intelligence based on authoritative data instead of heuristics. However, the Communities attribute lacks standardized values and semantics. Many operators document their Communities values in Internet Routing Registry (IRR) records, or their webpages, but typically not in machine-parsable format. To decipher the Communities values in an automated manner, we combine a web-mining tool with Google's Google Cloud Natural Language API~\cite{gcloud-nlp} to achieve the automatic compilation of a Communities dictionary, as explained in~\cite{giotsas2017detecting}.
As of March 2018, the dictionary included 11,830 interpreted Communities, 48\% of which encode geolocation data, 21\% encode relationship type, and the rest encode different types of routing policies (selective advertisement, blackholing, local preference tuning, path prepending).

Figure~\ref{fig:communities-methodology} illustrates the data extraction methodology. The system combines the interpreted Communities with a live stream of BGP data, obtained through through BGPStream~\cite{bgpstream}, to extract BGP updates annotated with the corresponding Communities. 
During the initialization phase, we continuously monitor the incoming BGP messages to establish a baseline of paths that are consistently tagged with a stable set of Communities. 
Then, CommunityWatch monitors the baseline of annotated paths to capture changes through explicit BGP withdrawals, or through changes in the attached Community values.
Routing updates are binned in time intervals to correlate path changes with routing incidents. The system uses a binning interval of 60 seconds (twice the default MRAI time~\cite{rfc4271}).
Whenever we detect a binning interval for which the paths deviating from the baseline exceed a minimum threshold, we trigger a signal of potential routing anomaly. Depending on the type of Communities, the corresponding signal investigation module analyzes the affected paths to determine the root cause of the observed change. 

\section{Anomaly Detection Use Cases}

The fact that BGP Communities encode different categories of routing meta-data, means that CommunityWatch can detect a wide range of routing anomalies. In this section we illustrate three such cases.

\textbf{Infrastructure Outages}
In the past, the AS-path data have been used to study changes in prefix
availability and reachability, and reveal the
occurrence of outages due to country-level censorship, attacks, or natural disasters~\cite{aceto2018comprehensive}. However, the coarse granularity of AS-paths has hindered detailed analysis of infrastructure outages, since many failures may change the infrastructure-level path, e.g., switching to another PoP, but the AS-path remains the same. Thus, an AS-path level and prefix-level analysis cannot show such failures.
Communities are often used to encode location information at fine granularities, such as IXP-level, and facility-level Points-of-Presence. Figure Figure~\ref{fig:franceix-outage-filtered} illustrates how using BGP Communities to filter the routing activity can reveal an outage at the France-IX IXP~\cite{franceix-outage}, by effectively de-noising the aggregated routing activity~\ref{fig:franceix-outage-unfiltered} that obscures the impact of localized events on the dynamics of BGP. We provide more details on the detection of infrastructure-level outages in~\cite{giotsas2017detecting}.

\textbf{Detection of Blackholed Prefixes}
Blackholing is a popular DDoS mitigation strategy inside a single
network or among multiple networks. BGP enables blackholing by leveraging the BGP communities attribute. Networks trigger blackholing requests by sending BGP announcements to their BGP neighbors for specific destination prefixes with the appropriate blackhole community.
Parsing of these values enable CommunityWatch to differentiate blackholing requests from normal BGP announcements, and characterize prefixes under attack, as we explain in~\cite{giotsas2017inferring}. Figure~\ref{fig:blackholing-growth} illustrates how CommunityWatch can be used to characterize the longitudinal growth of blackholing usage.

\textbf{Route leaks and policy violations}
Detection and analysis of export policy violations, such as violations of the valley-free rule, is of particular importance for
understudying BGP misconfiguration and characterizing misbehaving networks~\cite{rfc7908}. Detection of such violations requires accurate AS relationship data, but the universality of valley-free rule is a fundamental assumption of relationship inference algorithms. CommunityWatch detects such violations by parsing relationship-tagging communities which are free from inference heuristics biases. We analyzed BGP data in March 2018 to find that over 3\% of BGP paths violate the valley-free rule, which cannot be captured by the AS-Rank algorithm~\cite{CAIDA_Customer_cones_validation}.

\section{Acknowledgments}

The source code of CommunitiesWatch and detailed documentation is publicly available at~\url{https://github.com/vgiotsas/CommunitiesParser}.

\bibliographystyle{ACM-Reference-Format}
\bibliography{reference}

\end{document}